# A REVIEW ON ROUTING PROTOCOLS FOR APPLICATION IN WIRELESS SENSOR NETWORKS

Neha Rathi[1], Jyoti Saraswat[2] and Partha Pratim Bhattacharya[3]

Department of Electronics and Communication Engineering
Faculty of Engineering and Technology
Mody Institute of Technology & Science (Deemed University)
Lakshmangarh, Dist. Sikar, Rajasthan,
Pin – 332311, India

[1]`neharathi17@gmail.com`
[2]`jyotisaraswat.mit@gmail.com`
[3]`hereispartha@gmail.com`

## *ABSTRACT*

*Wireless sensor networks are harshly restricted by storage capacity, energy and computing power. So it is essential to design effective and energy aware protocol in order to enhance the network lifetime. In this paper, a review on routing protocol in WSNs is carried out which are classified as data-centric, hierarchical and location based depending on the network structure. Then some of the multipath routing protocols which are widely used in WSNs to improve network performance are also discussed. Advantages and disadvantages of each routing algorithm are discussed thereafter. Furthermore, this paper compares and summarizes the performances of routing protocols.*

## 1. INTRODUCTION

Due to recent progress in technology, there is a growth in wireless sensor network which comprises of large figure of homogeneous and heterogeneous sensor nodes which communicate in wireless fashion to achieve common objective. Homogeneous nodes are preferred over heterogeneous nodes because of less complexity and better manageability. Each sensor node communicates with other nodes within its radio communication range.

Nodes can be easily deployed in random or deterministic fashion and are normally battery operated. So, energy consumption is one of the most important factors. Wint Yi Poe[1], mainly focused on three competitors: uniform random, a square grid, and a pattern-based Tri-Hexagon Tiling(THT) node deployment under three performance matrices: coverage, energy consumption and worst–case delay which minimize the energy consumption, provide better coverage and guaranteed to extend the lifetime of the WSNs. In a class of three models, THT[1] defeats the other two models in terms of energy consumption and worst-case delay and square grid is preferred for better coverage performance. THT is well performing node deployment model for WSN applications. Eunil Park *et al.* in [2] proposed another method, a node scheduling method and a *protocol that considers both sides of Link Quality and Energy(PBLE),* an optimal routing protocol which is energy-efficient and prolong the lifetime of the sensor networks. PBLE [2], overcomes the problems arise in PRR× *Distance Greedy Forwarding* Method such as retransmission caused by loss of ACK transmission in real WSNs.

Usually, wireless sensor networks are composed of hundreds or thousands of sensor motes. Each node consists of processing capability (one or more microcontrollers, CPUs or DSP chips) and may hold several types of memory (program, data and flash memories), a RF transceiver (usually with a single omni-directional antenna), a power source (e.g., batteries and solar cells), and accommodate various sensors and actuators [3]. One or more nodes in the network will aid as sink(s) which exchange information with the user either directly or by the way of existing wired networks [4]. Peer-to-peer networking protocols support a mesh-like relation to switch data between the thousands of nodes in a multi-hop fashion. The flexible mesh architectures envisioned dynamically adapt to support introduction of new nodes or expand to cover a larger geographic region. Additionally, the system can automatically adapt to compensate for node failures [5].

The ideal wireless sensor is networked and scalable, fault tolerance, consume very little power, smart and software programmable, efficient, capable of fast data acquisition, reliable and accurate over long term, cost little to purchase and required no real maintenance.

The ways how to effectively route the collected dataamong nodes are the utmost important topic in WSNs because of the low powered sensor nodes. Based on the routing techniques and characteristics inWSNs, many routing protocols are proposed.The rest of the paper is structured as follows. Section 2 shows the main design constraints and routing challenges that routing protocols must face in wireless sensor networks. Section 3 describes the most popular routing protocols in wireless sensor networks. In section 4, comparisons of routing protocols are discussed. Section 5 draws the main conclusions of this work.

## 2.ROUTING CHALLENGES IN WIRELESS SENSOR NETWORKS

Due to reduced computing, radio and battery resources of sensors, routing protocols in wireless sensor network are expected to fulfill the following requirement:

**a) Data delivery model:**Data delivery model overcomes the problem of fault tolerance domain by providing the alternative path to save its data packets from nodes or link failures[6]**.** It severely affect the routing protocol in wireless sensor network, especially with regard to use the limited energy of the node, security purpose [7], energy consumption and route immobility.

**b) Scalability:**A system is said to be scalable if its effectiveness increases when the hardware is put-on and proportional to the capacity added [8]. Routing schemes make efforts with the vast collection of motes in WSNs which should be scalable enough to talk back to the events take place in the environment.

**c) Resilience**: Sometimes, due to environment problem or battery consumption sensors erratically stop working [9]. This problem is overcome by finding the alternate path when current-in use nodes stop operating.

**d) Production cost:**The cost of single node is enough to justify the overall cost of the sensor network. So the cost of each sensor node should be kept low.

**e) Operating environment:**Sensor network can be setup inside large machinery, at the base of the ocean, in a biologically or chemically contaminated field, in the battle field behind enemy line, in big building or warehouse etc.

**f) Power consumption:**Requirement such as long life time of sensor networks and restricted storage capacity of sensor nodes has directed to search a new scope to alleviate power consumption. Sidra Aslam discussed several schemes such as power aware protocol, cross-layer optimization, and harvesting technologies which help in reducing power consumption constraint in WSNs [10]. In multi-hop sensor networks, the multi-functioning of some nodes such as data sender and data router can cause topology change due to power failure which require new path for data transfer and restructure the network.

**g) Data aggression/fusion:** The main goal of data aggregation algorithms is to gather and aggregate data from different sources by using different functions such as suppression, min, max and average to achieve energy efficient and traffic optimization in routing protocols so thatnetwork lifetime is enhanced [11].

## 3. ROUTING PROTOCOLS IN WIRELESS SENSOR NETWORK

The sensor nodes are constrained to limited resources itself, so the main target is to design an effective and energy aware protocol in order to enhance the network lifetime for specific application environment. Since sensor nodes are not given a unified ID for identification and much redundant data collected at destination nodes. So, energy efficiency, scalability, latency, fault-tolerance, accuracy and QOS are some aspects which must be kept in mind while designing the routing protocols in wireless sensor networks.

Classically most routing protocols areclassified as data-centric, hierarchical and location based protocols depending on the network structure and applications. In data-centric routing, the sink sends queries to certain regions and waits for data from thesensors located in the selected regions. Since data isbeing requested through queries, attribute basednaming is necessary to specify the properties of data.Here data is usually transmitted from every sensornode within the deployment region with significantredundancy. Hierarchical or cluster based methods are well known techniques with special advantage of scalability and efficient communication. Nodes play different roles in the network. In location aware routing, nodes knowwhere they are in a geographical region. Location information is used to improve the performanceof routing and to provide new types of services.

Routing protocols canalso be classified into three categories such as proactive,reactive, and hybrid, depending on how thesource find a route to the destination. A proactive protocol sets up a routing paths and states before there is a demand for routing traffic. Paths are maintained even there is no traffic flow at that time. In reactive routing protocol, paths are set up on demand when queries are initiated. Hybrid protocol use combination of these two ideas.

A detail overview of routing protocols is discussed in the rest of the section.

### 3.1. Attribute-based or Data-centric Routing Protocols

The following protocols are discussed in this category:

#### 3.1.1. Flooding and Gossiping

Flooding and gossiping [12] are the most traditional network routing. They do not need to know the network topology or any routing algorithms. In flooding mechanism, each sensor receives a data packet and then broadcasts it to all neighboring nodes. When the packet arrives at the destination or the maximum number of hops is reached, the broadcasting process is stopped. On the other hand, gossiping is slightly enhanced version of flooding where the receiving node sends the packet to randomly selected neighbors, which pick another random neighbor to forward the packet to and so on.Although flooding is very easy, it has several drawbacks like implosion, overlap and resource blindness problem.

Gossiping avoid the problem of implosion by sending information to a random neighbor instead of classic broadcasting mechanism which send packets to all neighbors. However, gossiping creates another problem of delay in a propagation of data among sensor nodes.

### 3.1.2. SPIN

Joanna Kulik *et al.* in [12] proposed a family of adaptive protocol, called SPIN (Sensor Protocol for Information via Negotiation) that efficiently disseminate information among sensors in an energy-constrained wireless sensor network and overcome the problem of implosion and overlap occurred in classic flooding. Nodes running a SPIN communication protocol name their data using high-level data descriptors, called metadata. SPIN nodes negotiate with each other before transmitting data. Negotiation helps to ensure that the transmission of redundant data throughout the network is eliminated and only useful information will be transferred.

The SPIN family of protocols includes many protocols that disseminate information with low latency and conserve energy at the same time. The main two are called SPIN-1 and SPIN-2. Simulation result shows that SPIN-1 use negotiation to solve the difficulty of implosion and overlap. It reduces energy consumption by a factor of 3.5 when compared to flooding. As shown in figure 1 SPIN-2 is able to deliver even more data per unit energy than SPIN-1 and incorporate a threshold based resource-awareness mechanism in addition to negotiation, disseminates 60% more data per unit energy than flooding. Simulation result also shows that nodes with a higher degree tend to dissipate more energy than nodes with a lower degree, creating potential weak points in a battery-operated network.

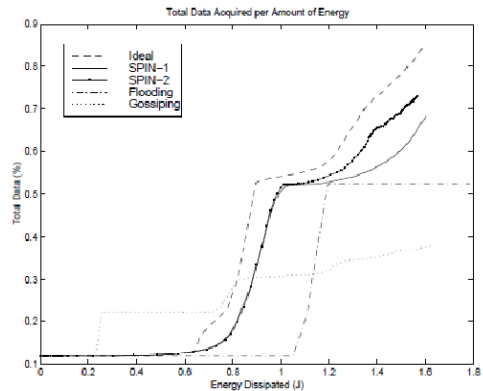

**Figure 1.** Data acquired for a given amount of energy. SPIN-2 distributes 10% more data per unit energy than SPIN-1 and 60% more data per unit energy than flooding.

The disadvantage of SPIN protocol is that it is not sure about the data will certainly reach the target or not and it is also not good for high-density distribution of nodes. Other drawback is that if the nodes that are interested in the data are far away from the source node and the nodes between source and destination are not interested in that data, such data will not be delivered to the destination at all. Therefore, SPIN is not a good choice for applications.

### 3.1.3. Directed Diffusion

Ramesh Govindan*et al.* in [13] proposed a popular data aggregation paradigm for wireless sensor networks called directed diffusion. Directed diffusion is data-centric and all nodes in a directed diffusion-based network are application-aware. This enables diffusion to achieve energy savings by selecting empirically good paths and by caching and processing data in-network (*e.g.,* data aggregation).

Directed diffusion is composed of several elements: interests, data messages, gradients, and reinforcements as shown in figure 2. An *interest* message is a query which specifies what a user wants and containsa summary of a sensing task which is supported by a sensor network for acquiring data. Typically, *data* in sensor networks is the collected data of a physical phenomenon. Such data can be an *event* which is a short description of the sensed phenomenon. In directed diffusion, data is *named* using attribute-value pairs.

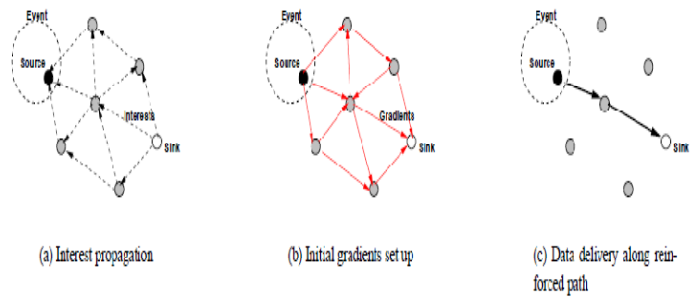

**Figure 2.** A simplified schematic for directed diffusion.

SPIN protocol allow sensors to advertise the availability of data and the nodes which are interested query that data but in Directed Diffusion the sink query the sensor nodes if a specific data is available by flooding.

The main advantages of directed diffusion are:

1) Since it is data centric, communication is neighbor-to-neighbor with no need for a node addressing mechanism. Each node can do aggregation and caching, in addition to sensing. Caching is a big advantage in term of energy efficiency and delay.

2) Direct Diffusion is highly energy efficient since it is on demand and there is no need for maintaining global network topology.

Directed Diffusion is not a good choice for the application such as environmental monitoring because it require continuous data delivery to the sink will not work efficiently with a query-driven on demand data model.

### 3.1.4. Rumor Routing

Rumor routing is proposed in [14], which allows queries to be delivered to events in the network. It is mainly determined for context in which geographic routing criteria is not applicable. Rumor routing is a logical compromise between flooding queries and flooding events notification.

Rumor routing is tunable and allows for tradeoff between setup overhead and delivery reliability. Generally, directed diffusion floods the queries to the entire network and data can be sent through multiple paths at lower rates but rumor routing maintains only one path between source and destination as shown in figure 3. In this protocol, paths are created for queries to be delivered and when a query is generated it is sent for random walk until it finds the path, instead of flooding it throughout the network. When the event path is discovered by the query, it can be routed directly to the event.

When events are flooded through the network, node detects an event, maintains its event table and creates an agent. The table entries contain the information about source node, events and last hop node. The main job of the agent is to propagate the information about local events to distant nodes.

Simulation result shows that rumor routing protocol is reliable in terms of delivering queries to events in large network, handle the node failure very smoothly and degrading its delivery rate linearly with the number of failure nodes. It also achieves significant energy saving over event flooding.

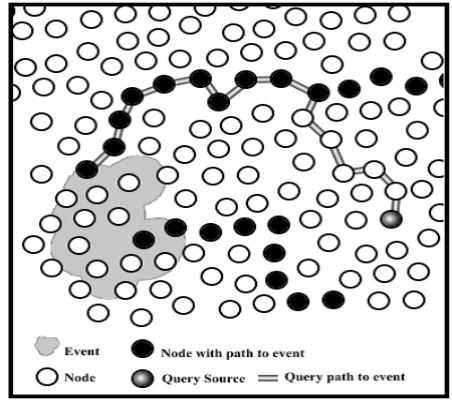

**Figure 3.** Query is originated from the query source andsearch for a path to the event. As soon as it finds a node onthe path, it's routed directly to the event.

### 3.1.5. Gradient-Based Routing

The algorithm makes an improvement on Directed Diffusion, in order to get the total minimum hop other than the total shortest time. In the traditional gradient minimum hop count algorithm, hop count is the only metric, which measures the quality of route. Li Xia in [15] proposed a new gradient routing protocol which not only consider the hop count but also use the remaining energy of each node while relaying data from source node to the sink. This scheme is helpful in handling the frequently change of the topology of the network due to node failure.

A new gradient routing scheme also aims to establish the cost field and find a minimum cost path from the source node to the sink. Figure 4illustrates a simple example of the procedure of generating minimum cost gradient.

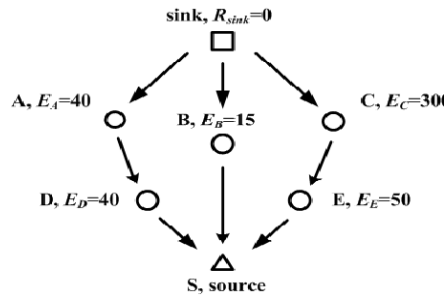

**Figure 4.**The procedure of generating minimum cost gradient.

The source node has three routes to reach the sink, route1: S–>D–>A, rout2: S–>B and route3:S–>E–>C. At the routing setup stage, the source node will receive three different setup messages. The cost metric of route1 is: 1/40+1/40=1/20. It is the smallest cost in these threeroutes. So the source node will choose route1 as its optimal route. It records node *D* as its previous relay node. After a period, the nodes in route1 may have low energy level. In this situation, route2 may be chosen as the good route.

In the route setup stage, when one node receives the setup message, it waits for a short time $T_{wait}$for messages with better metric, which may arrive during this period. When $T_{wait}$expires, the node rebroadcasts the message with the best metrics in all messages it has received. By this way, the number of setup messages in the whole network can decreasegreatly. This scheme aggregates similar packets into one packet and transmits the information, which is energy efficient and helps to prolong the network's lifetime.

44

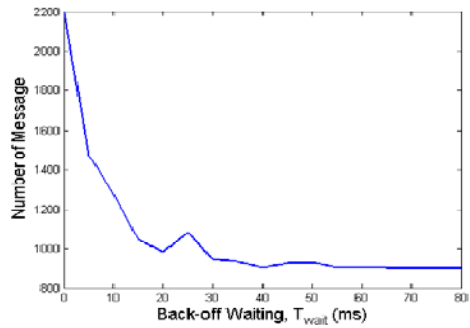

**Figure 5.** The relationship between the number of setup messages and $T_{wait}$.

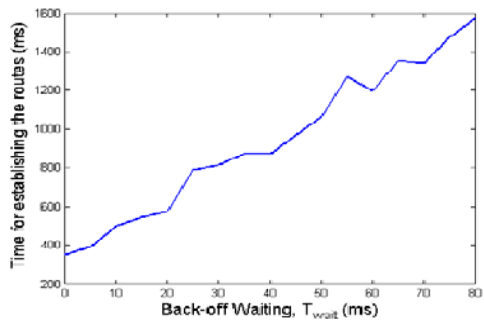

**Figure 6.** The relationship between routes establishment time and $T_{wait}$.

Simulation result shows the relationship between the number of relayed setup messages and the back-off waiting time $T_{wait}$ as shown in figure 5. When $T_{wait}$ is large, the total messages for setting up the network's route will be small in number. So the back-off waiting scheme is quite effective for saving the energy consumption when establishing the network's routes. However, it delays the establishment of routes for a while. The size of such delay is basically proportional to $T_{wait}$, as illustrated in figure 6.

### 3.2. Hierarchical-Based Routing(Clustering)

Hierarchical routing is a guarantee approach for point-to- point routing with very small routing state [16]. Scalability is one of the essential design features of the sensor networks. Single-gateway architecture can cause the gateway to overload which might cause a break in communication and tracking of events is unhealthy. Other major disadvantage is that long –haul communication is not possible because it is not scalable for large set of sensors. To overcome these drawbacks network clustering has been pursued in some routing approaches.

Hierarchical or cluster based methods are well known techniques with special advantage of scalability and efficient communication. Nodes play different roles in the network. Hierarchical routing maintains the energy consumption of sensor nodes and performs data aggregation which helps in decreasing the number of transmitted messages to base station. The whole WSN is divided into a number of clusters in term with the specific rules.Some hierarchical protocols are discussed here.

#### 3.2.1. LEACH

LEACH [17] stands for Low-Energy Adaptive Clustering Hierarchy and is one of the first hierarchical protocols. When the node in the network fails or its battery stops working then LEACH protocol is used in the network. Leach is self-organizing, adaptive clustering protocol in which sensor nodes will organize themselves into local clusters and cluster members elect cluster head (CH) to avoid excessive energy consumption and incorporate data aggregation which reduces the amount of messages sent to the base station, to increase the lifetime of the network.Therefore this algorithm has an effect on energy saving.

Cluster head is responsible for collecting data from its cluster members. To reduce intercluster and intracluster collisions, LEACH uses a TDMA/code-division multiple access(CDMA). The decision whether a node elevates to cluster head is made dynamically at a time interval. However, data collection is performed periodically. Therefore, the LEACH protocol is mainly used for constant tracking by the sensor networks. When the node becomes cluster head for the current round, then each elected cluster head broadcasts information to rest of the nodes in the network.

To balance the energy dissipation of nodes, cluster heads change randomly over time [18]. The node makes this decision by choosing a random number between 0 and 1. The node becomes cluster head for the current round if the number is less than the following threshold:

$$T(n) = \frac{p}{1 - p(r \bmod (1/p))} \text{ if } n \in G,$$

Where n is the given node, P is theapriori probability of a node being elected as a cluster head, r is the current round number and G is the set of nodes that have not been elected as cluster head in the last 1/P rounds.

|  | SPIN | LEACH | Directed Diffusion |
|---|---|---|---|
| Optimal Routing | No | No | Yes |
| Network Lifetime | Good | Very Good | Good |
| Resource Awareness | Yes | Yes | Yes |
| Use of Meta-Data | Yes | No | Yes |

**Table1.** Comparison between SPIN, LEACH, and Directed Diffusion.

Table 1shows the comparison between SPIN, LEACH and Directed Diffusion according to different parameters. A centralized version of this protocol is Leach-C. This scheme is divided into two phases, the set-up phase and the steady-phase. In set-up phase, sensors communicate with the base station and tell it about their current position and about their energy level. In steady-state phase, the actual data transfer to the base station takes place. The duration of steady-phase is longer than the set-up phase to minimize overhead.

Two-Level Hierarchy LEACH (TL-LEACH) is a modified form of the LEACH algorithm which consists of two levels of cluster heads (primary and secondary) instead of a single one. The advantage of two-level structure of TL-LEACH is that it reduces the amount of nodes that transmit information to the base station, effectively reducing the total energy usage.

### 3.2.2. PEGASIS and Hierarchical-PEGASIS

PEGASIS (Power-EfficientGathering in Sensor Information Systems), a near optimal chain-based protocol that is an improvement over LEACH. Instead of forming multiple clusters, PEAGSIS construct a node chain when nodes are placed randomly in a play field then each node communicates only with a close neighbor and takes turns transmitting to the basestation, thus reducing the amount of energy spent perround [19]. The chain construction is performed in a greedy way. Figure 7 shows node 0 connected node 3, node 3 connecting to node 1, and node 1 connecting to node 2. When a node fails, the chain is reconstructed in the same manner by avoiding the dead node.

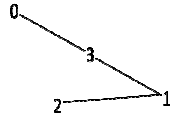

**Figure 7.** Chain construction using the greedy algorithm.

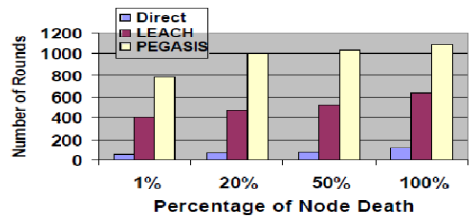

**Figure 8.** Performance results for 50m x 50m network with initial energy 0.25J/node.

Simulation result in figure 8 shows that PEGASIS performs better than LEACH by about 100 to 300%when 1%, 20%, 50%, and 100% of nodes die for differentnetwork sizes and topologies.

Hierarchical-PEGASIS conducts a further improvement; it allows concurrent transmission when the nodes are not adjacent. Compared with LEACH, the two algorithms eliminate the overhead of forming cluster, but both of them do not take the energy condition of next hop into consideration when choosing a routing path, so they are not suitable for heavy-loaded network. When the amount of nodes is very large in WSNs, the delay of data transmission is very obvious, so they do not scale well and also are not suitable for sensor networks where such global knowledge is not easy to obtain.

### 3.2.3. TEEN and APTEEN

TEEN stands forThreshold sensitive Energy Efficient sensorNetwork protocol. It is the first protocol developed for reactive networks and used in temperature sensing application[20]. Based on LEACH, TEEN is based on hierarchical grouping which divides sensor nodes twice for grouping cluster in order to detect the scene of sudden changes in the sensed attributes such as temperature. After the clusters are formed, TEEN separates the Cluster Head into the second-level Cluster Head and uses Hard-threshold and Soft-threshold to detect the sudden changes.The model is depicted in figure 9.

Thus, the hard threshold tries to reduce the number of transmissions by allowing the nodes to transmit only when the sensed attribute is in the range of interest. The soft threshold further reduces the number of transmissions by eliminating all the transmissions which might have otherwise occurred when there is little or no change in the sensed attribute once the hard threshold.

The main drawback of this scheme is that it is not well suited for applications where the user needs to get data on a regular basis. Another possible problem with this scheme is that a practical implementation would have to ensure that there are no collisions in the cluster. TDMA scheduling of the nodes can be used to avoid this problem but this causes a delay in the reporting of the time-critical data. CDMA is another possible solution to this problem. This protocol is best suited for time critical applications such as intrusion detection, explosion detection etc.

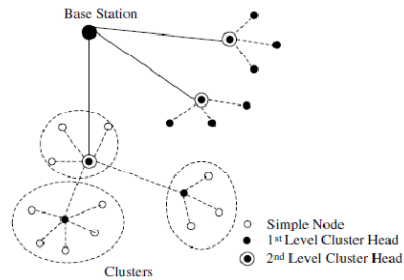

**Figure 9.** Hierarchical clustering in TEEN and APTEEN.

The Adaptive Threshold Sensitive Energy Efficient Sensor Network Protocol (APTEEN) is an extension of TEEN and aims at both capturing periodic data collections and reacting to time critical events. The architecture is same as in TEEN. In APTEEN once the CHs are decided, in each cluster period, the cluster head broadcasts the parameter such as attributes, threshold,schedule and count time to all nodes [21].

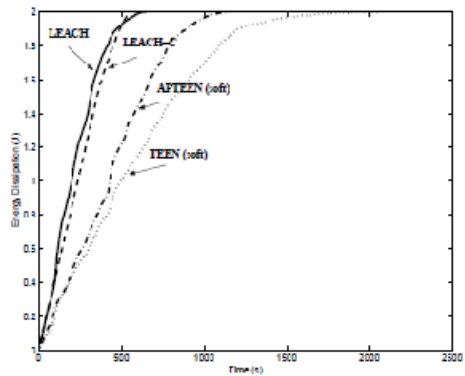

**Figure 10.** Comparison of average energy dissipationfor LEACH, APTEEN and TEEN.

Simulation result compare APTEEN with TEEN and LEACH(*leach and leach-c)* with respect to energy consumptionas shown in figure 10. The performance of APTEEN lies between TEEN and LEACH in terms of energy consumption and longevity of the network. While sensing the environment, TEEN only transmits time critical data.APTEEN makes an improvement over TEEN by supporting periodic report for time-critical events. The main disadvantages of the two algorithms are the overhead and complexity of forming clusters.

**3.2.4. Energy-aware cluster-based routingalgorithm**

Jyh-Huei Chang *et al.* in [22] proposed Energy-Aware, Cluster-Based Routing Algorithm (ECRA) for wireless sensor networks to maximize the network's lifetime. The ECRA selects some nodes as a cluster-heads to construct Voronoi diagrams and cluster-head is rotated to balance the load in each cluster.

LEACH may have several problems: First, if the coverage of the cluster-heads is too small, then some cluster-heads may not have any members in their clusters. Second, LEACH has a long transmission range between the cluster-heads and the sink node. Third, the LEACH requires global cluster-heads rotation. This cluster-head selection greatly increases processing and communication overhead, thereby consuming more energy. Therefore, this protocol is used to overcome the LEACH's problems and reduce the overhead of cluster-heads rotation for cluster-based wireless sensor networks. Clustering, data transmission and intra-cluster-head rotation are the phases of energy aware cluster based routing protocol.

It is shown in figure 11that the cluster members of the cluster transmit their sensing data to cluster-heads which forward the aggregated data to the sink node. ECRA helps in balancing the load for all sensors and avoid too many cluster-heads focusing on a small area by choosing a sensor node from the previous cluster as a cluster-header, called an intra-cluster-head rotation.

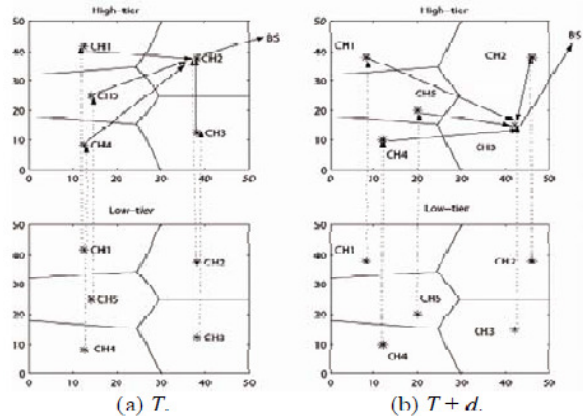

**Figure 11.** The operation of high-tier architecture in enhanced ECRA, where $T$ is the current round and $T + d$ is the next round, and so on.

ECRA-2T is two-tier architecture for ECRA which is used to enhance the performance of the original ECRA. The ECRA can be advanced by adding an extra tier called a high tier. The high tier has only one cluster and all cluster-heads inthe low tier arealso the members in the high tier. The nodes in the high-tier forward their aggregated data to the node with the maximal remaining energy, called the *main cluster-head*.

The main cluster-head transmits the aggregated data to the sink. When a round isover, rotate the cluster-head of the low-tier in the sensing field based on the parameter $O_{ij}$:

$$O_{ij} = \frac{E_{ij}^{new}}{E_{d_{ij}}}, i = 1, \ldots, n, j = 1, \ldots, |C_i|.$$

The members of the high-tier in the next round consist of these cluster heads. In current round $T$, $CH2$ is the main cluster-head. In the next round, $T + d$, $CH3$ has a maximal remaining energy that is selected as the main cluster-head, and so on.

The simulation result shows that both ECRA-2T and ECRA outperform all other routing schemes: direct communication, static clustering, and LEACH. The system lifetime of ECRA-2T is approximately 2.5 times than that of LEACH. ECRA-2T also requires much less energy consumption than that of direct communication as shown in figure 12.

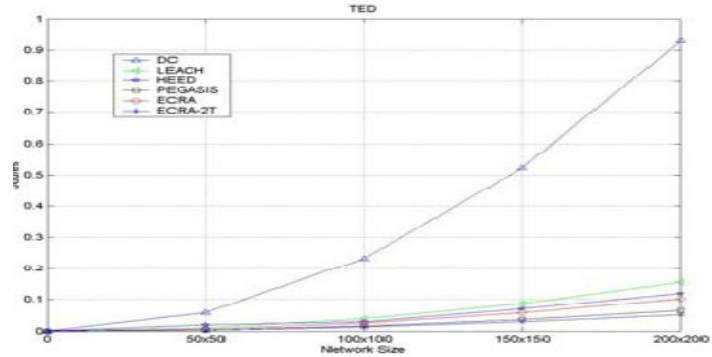

Figure 12. Total energy dissipation (TED) usingdirect communication, LEACH, ECRA,and ECRA-2T. The messages are 2,000bits.

### 3.3. Location-Based Routing (Geographic Protocol)

Most of the routing protocols require location information for sensor nodes in wireless sensor networks to calculate the distance between two particular nodes on the basis of signal strength so that energy consumption can be estimated. It is also utilized in routing data in energy efficient way when addressing scheme for sensor network is not known. It is worth noting that there have been many location-based protocols in Ad Hoc networks and it makes great effects when we transplant those research achievements for wireless sensor networks in some ways.

#### 3.3.1. MECN and SMECN

Minimum EnergyCommunication Network (MECN) [23] sets up and maintains a minimum energy network for wireless networks by utilizing low power GPS. Although, the protocol assumes a mobile network, it is best applicable to sensor networks, which are not mobile. A minimum power topology for stationary nodes including a master node is found. MECN assumes a master site as the information sink, which is always the case for sensor networks.

MECN identifies a relay region for every node. The relay region consists of nodes in a surrounding area where transmitting through those nodes is more energy efficient than direct transmission. The relay region for node pair (*i, r*) is depicted in Figure 13.

The enclosure of a node *i*is then created by taking the union of all relay regions that node *i* can reach. The main idea of MECN is to find a sub-network, which will have less number of nodes and require less power for transmission between any two particular nodes.In this way, global minimum power paths are found without considering all the nodes in the network. This is performed using a localized search for each node considering its relay region.

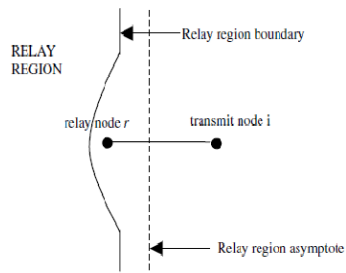

**Figure 13.**Relay Region of transmit relay node pair (*i, r*) in MECN.

MECN is self-reconfiguring and thus candynamically adapt to nodes failure or thedeployment of new sensors. Between two successivewake-ups of the nodes, each node can executethe first phase of the algorithm and the minimumcost links are updated by considering leaving ornewly joining nodes.

The Small Minimum Energy CommunicationNetwork (SMECN) [24] is an extension to MECN.In MECN, it is assumed that every node cantransmit to every other node, which is not possibleevery time. In SMECN possible obstacles betweenany pair of nodes are considered. However, thenetwork is still assumed to be fully connected as inthe case of MECN. The sub-networkconstructedby SMECN for minimum energy relaying isprovably smaller (in terms of number of edges)than the one constructed in MECN if broadcastsare able to reach to all nodes in a circular regionaround the broadcaster. As a result, the number ofhops for transmissions will decrease. Simulationresults show that SMECN uses less energy than MECN and maintenance cost of the links is less. However, finding a sub-network with smaller number of edges introduces more overhead in the algorithm.

### 3.3.2. GEAR (Geographic and Energy Aware Routing)

The aim is to reduce the number of Interest in DirectedDiffusion and add geographic information into interest packetby only considering a certain region rather than sendingInterest to the whole network by means of flooding. GEARuses energy aware and geographically informed neighborselection heuristics to route a packet towards the target region [25].Therefore, GEAR helps in balancing energy consumption in thisway and increase the network lifetime. When a closer neighbor to the destination exists, GEAR forwards the packet to the destination by picking a next-hop among all neighbors that are closer to the destination. When all neighbors are far away, there is a hole then GEAR forward the packet by picking a next-hop node that minimizes some cost value of this neighbor. Recursive Geographic Forwarding algorithm is used to disseminate the packet within the region.

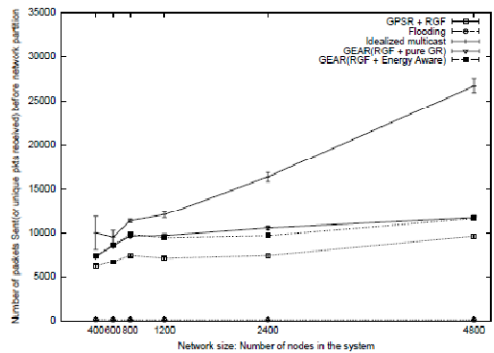

**Figure 14.** Comparison for uniform traffic.

GEAR is compared to a similar non-energy awarerouting protocol GPSR, which is one ofthe earlier works in geographic routing that usesplanar graphs to solve the problem of holes. The simulation results show that for uneven traffic distributions, GEAR delivers 70% to 80% morepackets than GPSR. For uniform traffic pairs, GEAR delivers25 - 35% more packets than GPSR as shown in figure 14.

### 3.3.3. GAF and HGAF

GAF (Geographic Adaptive Fidelity) [26] is adaptive fidelity algorithm in which large numbers of sensor nodes are placed in observed area and only few nodes in the observed area are selected to transmit messages, while the other nodes sleep. In this way, GAF reduces the number of nodes needed to form a network and saves nodes' battery.

Hierarchical Geographical Adaptive Fidelity (HGAF) saves much more battery by enlarging the cell of GAF by adding a layered structure for selecting an active node in each cell. GAF saves battery power by enlarging the size of the cell. The connectivity between active nodes in two adjacent cells must be guaranteed because active nodes works as cluster heads to deliver packets between cells. Because of this limitation, GAF needs an active node in every area whose maximum size is $R^2/5$.

HGAF limits the position of active node in a cell and synchronizes the position in each cell among all cells. Through this modification, the connectivity between active nodes in two adjacent cells can be guaranteed for a larger cell than in GAF.

Simulation result shows that HGAF outperforms GAF in terms of survived nodes and the packet delivery ratio when the node density is high. The lifetime of dense and randomly distributed networks with HGAF is about 200% as long as ones with GAF.

### 3.3.4. Fermat Point Based Energy Efficient Geocast Routing Protocol

Geocast routing protocol is used to deliver packets to a group of nodes that are within a specified geographical area, i.e., the geocast region. Fermat point based protocols are adapted for reducing the energy consumption of a WASN by reducing the total transmission distance in a multi hop-multi sink scenario. Congested environment around a WASN expand the chance of multipath propagation and it in turn acquaint multipath fading. In [27], the effects of both of these factors are considered on the performance of I-Min routing protocol designed for WASNs. I-MIN is the energy efficient scheme as it increases the probability that a node with higher residual energy is selected even if its distance from destination is somewhat more as compared to that for another node with a lesser value for residual energy.

After modifying the radio model with considerations for changed propagation environmental effects and multipath fading, the consumption of energy in a geocast routing protocol is shown to vary considerably. Higher the number of geocast regions, larger is the total distance that a data packet has to travel and thereby greater is the effect of propagation environment combined with the effect of multipath fading on the performance of an energy aware algorithm.

### 3.4. Multipath Routing Protocol

Due to the limited capacity of a multi-hop path [28]and the high dynamics of wireless links, single-path routing approach is unable to provide efficient high data rate transmission in wireless sensor networks. Nowadays, the multipath routing approach is broadly utilized as one of the possible solutions to cope with this limitation. This section discusses some of the multipath routing protocols.

### 3.4.1. N-to-1 Multipath Routing Protocol

N-to-1 Multipath Routing protocol [29] is proposed according to the converge cast traffic pattern of wireless sensor networks. In this technique, a multiple node disjoint paths are simultaneously discovered from all sensor nodes towards a single sink node. In this protocol, the sink node discover route by sending a *route update* message and this stage called *branch-aware flooding*, which discover several paths from sensor node towards a single sink tree and construct a spanning tree. Then each sensor node that receives a *route update* message for the first time, selects the sender of this message as its parent towards the sink node. In addition, if an intermediate node overhears a *route update* message from another neighboring node that introduces an alternative node-disjoint path through a different branch of the spanning tree, it adds this path to its routing table.This process continues until all sensor nodes discover their primary path towards the sink node and spanning tree is constructed through all the nodes as shown in figure 6(a). After that,*multipath extension flooding technique* is used to discover more paths from each sensor node towards the sink node. As shown in figure 15(b), each link between two individual nodes that belong to different branches of the constructed spanning tree can help to establish an additional path from these nodes towards the sink node.

This protocol utilizes the single-path forwarding strategy for transmitting each data segment, while all the intermediate nodes use an adaptive per-hop packet salvaging technique to provide fast data recovery from node or link failures along the active paths.

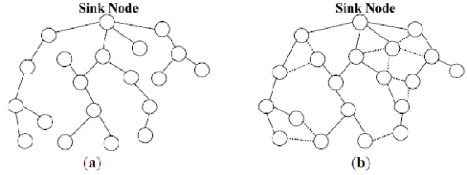

**Figure 15.** (**a**) Spanning tree constructed by initial flooding in N-to-1 Multipath Routing Protocol. (**b**) Multipath discovery using multipath extension flooding mechanism.

N-to-1 multipath routing protocol profits from the availability of several paths at the intermediate nodes toimprove reliability of packet delivery by employing a per-hop packet salvaging strategy. Nevertheless,using such a simple flooding strategy cannot result in constructing high-quality paths with minimuminterference. According to the operation of this protocol, concurrent data transmission over constructed paths may reduce thenetwork performance.

### 3.4.2. Multipath Multispeed Protocol (MMSPEED)

MMSPEED [30] is designed based on the cross-layer design approach between network and MAC layer to provide QoS differentiation in terms of reliability and timeliness.MMSPEED is the extension of the SPEED protocol [31] which guarantees timeliness packet delivery by introducing multiple speed levels and provide different speed layers over a single network. In this protocol, data packets are assigned to the appropriatespeed layer to be placed in the suitable queue according to their speed category. After that, datapackets are serviced in the FCFS policy.

Fromreliability perspective, MMSPEED benefits from path diversity property of multipath routing approachto guarantee reliability requirements of each data packet. This protocol provides reliability differentiationthrough controlling number of active paths and sending multiple copies of the original data packetsover several paths. Accordingly, each intermediate node selects a set of next-hop neighboring nodestowards the destination node based on the estimated packet loss rate over each link and theirgeographic distance from itself.

MMSPEED satisfy the delay requirements of various applications. To satisfydifferent delay requirements, each intermediate node tries to forward its received data packet to theneighboring node, which is closer to the destination node in order to provide a good speed progress.However, according to the experimental results provided in [32], probability of successful datatransmission over low-power wireless links highly depends on the sender-receiver distance andinterference power of the receiver. Therefore, using geographic routing with greedy forwarding doesnot necessarily improve network performance metrics. Moreover, since data transmission over long links exacerbates the required energy for data transmission; this protocol cannot support long-lifeapplications.

### 3.4.3. Braided Multipath Routing Protocol

Braided Multipath Routing Protocol [33] is a seminal multipath routing protocol proposed toprovide fault-tolerant routing in wireless sensor networks. This protocol uses a same technique as inDirected Diffusion, uses two types of path *reinforcement* messages to construct several partially disjoint paths.

The sink node sends a *primary path reinforcement* message to its best neighboring node to initiate the path. When an intermediate node receives a *primary path reinforcement* message, it forwards this message to its best next-hop neighboring nodetowards the source node. This process is repeated until the *primary path reinforcement* messagereaches the source node. Whenever the sink and intermediate nodes send out the *primary pathreinforcement* message, they also generate an *alternative path reinforcement* message and send thismessage to their next preferred neighboring node towards the source node. Through establishing a set of partially disjoint paths betweenthe source and sink nodes, whenever the primary path fails to forward data packets towards the sinknode, one of the constructed alternative paths can be utilized to avoid data transmission failure.

Simulation result shows the comparison of the lower overhead of braided multipath routing with the idealized node-disjoint multipath protocol.The proposedapproach provides about 50% higher resilience against path failures, compared to the idealized node disjoint multipath protocol. Besides, since this approach isdesigned based on the principles of Directed Diffusion, the drawbacks of Directed Diffusion can bepresent in this protocol.

### 3.4.4. Energy-Aware Routing

Energy aware routing protocol is efficient method to minimize the energy cost for communication and can increase the network lifetime. Unlike directed diffusion, data transmission is done through several optimum paths at higher rates instead of transmitting through one optimal path. The transmission path selection is done by choosing a probability value of each path. The probability value balanced the initial network load and enhanced the network lifetime.

An energy aware routing protocol is proposed in [34] which provide a reliable transmission environment with low energy consumption. It is used for making decision on which neighbor a sensor node should forward the data message. A node is selected on the basis of its residual energy level and signal strength. Ideally, the greater the energy of the node is more likely to be selected on the next hop. The nodes which are not selected will move to the sleep state to conserve power. Network connectivity is shown in figure 16 [34]. There are many intermediate nodes available in the network. All nodes within the radio range of the nodes receive the broadcast message at the same time. When the sink initially broadcast the message, the nodes A, E and G receive the message. Assume that the available energy at A is larger than at E and G, and also A is within the required signal strength threshold, hence node A is selected to broadcast the message to the neighboring nodes. The process continues and node B which is selected sends out the broadcast message which is received by nodes F and C, it is found that both F and C have the same energy level and are within the required signal strength threshold.

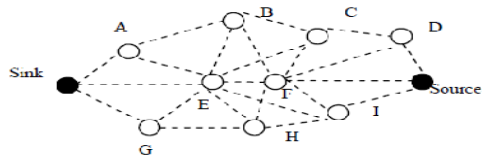

**Figure 16.**Network connectivity.

So both F and C start a back-off timer and if the back-off timer of node F ends before C an implicit acknowledgement is sent by node F which is also received by node C, and so node C stops its back-off timer as shown in Figure 17. When the broadcast message reaches the target source, the source transmits the route reply packet through the nodes it received the broadcast message. This protocol provides reliable packet delivery for unicast transmission. Data is cached in the sender until an ACK is received from the receiver. If no ACK is received within a timeout period, an error report is generated and the data will be sent back to the original source of this data in order to retransmit.

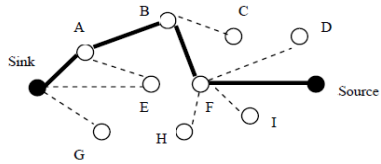

**Figure 17.** Path selected in energy aware routing.

The disadvantage is that energy-aware routing needs to exchange local information between neighbor nodes and all nodes have a unified address, which enlarges the price of building routing paths.

## 4. COMPARISON OF ROUTING ROTOCOLS

In Table 2 the above routing protocols are compared according to their design characteristics.

| Routing protocols | Classification | Power usage | Data-aggregation | Multipath | Query-based | QoS |
|---|---|---|---|---|---|---|
| SPIN | Flat | Ltd. | Yes | Yes | Yes | No |
| Directed diffusion | Flat | Ltd. | Yes | Yes | Yes | No |
| Rumor routing | Flat | Low | Yes | No | Yes | No |
| GBR | Flat | Low | Yes | No | Yes | No |
| LEACH | Hierarchical | High | Yes | No | No | No |
| PEGASIS | Hierarchical | Max. | No | No | No | No |
| TEEN and APTEEN | Hierarchical | High | Yes | No | No | No |
| ECRA | Hierarchical | Max. | Yes | No | No | No |
| MECN and SMECN | Hierarchical | Low | No | No | No | No |
| GEAR | Location | Ltd. | No | No | No | No |
| GAF | Location | Ltd. | No | No | No | No |
| N-to-1 multipath | Flat | Ltd. | Yes | Yes | No | No |
| MMSPEED | QoS | Low | No | Yes | No | Yes |
| Braided multipath | Flat | Ltd. | Yes | Yes | Yes | No |
| Energy aware | Flat | N/A | No | No | Yes | No |

**Table 2.** Classification and comparison of routing protocols in WSN.

## 5. CONCLUSION

Routing in sensor networks is a new research area, with a limited but rapidly growing set of results. In this paper, routing protocols are discussed based on three categories: Flat based routing, Hierarchical-based routing and Location-based routing on the basis of network structure. They have the common objective of trying to extend the lifetime of the sensor network. Rumor routing discussed in the paper is tunable and allows for tradeoff between setup overhead and delivery reliability. In Gradient Based routing the back-off waiting scheme is quite effective for saving the energy consumption when establishing the network's routes. Hierarchical based techniques have special advantage of scalability and efficient communication. Hierarchical routing maintains the energy consumption of sensor nodes and performs data aggregation which helps in decreasing the number of transmitted messages to base station. Most of the routing protocols require location information for sensor nodes in wireless sensor networks to calculate the distance between two particular nodes on the basis of signal strength so that energy consumption can be estimated. Single-path routing approach is unable to provide efficient high data rate transmission in wireless sensor networks due to the limited capacity of a multi-hop path and the high dynamics of wireless links. This problem can be overcome by using multipath routing. Many routing protocols have been proposed which are not suitable for all applications in WSNs. Many issues and challenges still exist that need to be solved in the sensor networks.

**Authors**

NehaRathi was born in India on 16 July,1987. She received her B. Tech degree in Electronics and Communication Engineering from Uttar Pradesh Technical University, India in 2010 and currently is a M. Tech (Signal Processing) student in Mody Institute of Technology and Science (Deemed University), Rajasthan, India.

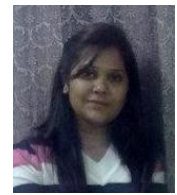

JyotiSaraswat was born in India on November 26,1987. She received herB.Tech degree in Electronics and Communication Engineering from Modi Instituteof Engineering and  Technology, Rajasthan University, India in 2010and currently is a M. Tech (Signal Processing) studentin ModyInstitute of Technology and Science (Deemed University), Rajasthan, India.

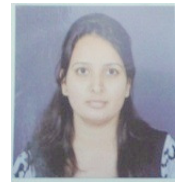

Dr. ParthaPratim Bhattacharya was born in India on January 3, 1971. He has 15 years of experience in teaching and research. He served many reputed educational Institutes in India in various positions starting from Lecturer to Professor and Principal. At present he is working as Professor in Department of Electronics and Communication Engineering in the Faculty of Engineering and Technology, Mody Institute of Technology and Science (Deemed University), Rajasthan, India. He worked on Microwave devices and systems and mobile cellular communication systems. He has published a good number of papers in refereed journals and conferences. His present research interest includes wireless sensor networks, mobile communication and cognitive radio.

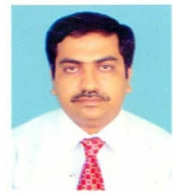

Dr. Bhattacharya is a member of The Institution of Electronics and TelecommunicationEngineers, India and The Institution of Engineers, India. He is the recipient of Young Scientist Awardfrom International Union of Radio Science in 2005. He is working as the chief editor, editorial boardmember and reviewer in many reputed journals.